# Low Density Parity Check Code (LDPC Codes) Overview

Saumya Borwankar
Department of Electronics and Communication Engineering
Institute of technology, Nirma University
17bec095@nirmauni.ac.in

Dhruv Shah
Department of Electronics and Communication Engineering
Institute of technology, Nirma University
17bec097@nirmauni.ac.in

*Abstract—* **This paper basically expresses the core fundamentals and brief overview of the research of R. G. GALLAGER [1] on Low-Density Parity-Check (LDPC) codes and various parameters related to LDPC codes like, encoding and decoding of LDPC codes, code rate, parity check matrix, tanner graph. We also discuss advantages and applications as well as the usage of LDPC codes in 5G technology. We have simulated encoding and decoding of LDPC codes and have acquired results in terms of BER vs SNR graph in MATLAB software. This report was submitted as an assignment in Nirma University**

## I. INTRODUCTION

With the continuous evolution of digital wireless communication and information technology, an efficient communication system in terms of bandwidth, speed, cost, lossless data transmission and secure communication is must for any advancements. We do not have any control over the channel and hence the effects in terms of losses happening due to noise in the channel on wireless data transmission are not liable to variation. So, that is where an efficient channel coding scheme comes which should be able to detect and correct any amount of errors at the receiver which are caused due to noise in wireless channel so that we can retrieve the original data and thus resulting in a successful wireless transmission. But as the generations in information technology advance and the number of data in terms of bits increases, various parameters like complexity of structure for implementation of such logic of channel coding, encoding and decoding speed or we can say computation time, cost of implementation are needed to be improvised. So, that's where the implementation of LDPC codes can fulfil these different parameters for a quite increased efficiency.

LDPC codes are very efficient codes since they offer practical implementation at nearly achieving the Shannon channel capacity of reliable transmission. Shannon channel capacity rule denotes that the code with the code rate near to the capacity number gives the error going to zero in decoding with the maximum likelihood decoder as we increase the block length. [2] This condition can be achieved by using random linear block codes that are encoded as polynomial of time. But as we increase the block length for error approaching to zero, there comes the issue of complex computation algorithms for encoding and decoding and that eventually leads us to the compromise with the speed. So, here LDPC code fulfils the successful implementation of LDPC code at Shannon limit with long block lengths and also having additional advantages of less complex algorithms and greater speed and accuracy. So how do LDPC codes fulfil these conditions that we will briefly understand in this paper. So, before we go to understand the LDPC code, we need to understand some concepts of linear block code and the code rate.

### A. Linear Block Codes

In Linear Block Code, parity bits and message bits are in linear combination and this combination together form a code word. Parity bits in transmitted codewords over a noisy channel help us at the receiver side by detecting, correcting and locating the errors with the help of decoding algorithms. This Linear codeword can be represented as (n,k) codeword where, n represents the block length of encoded code and k represents the message bit.

### B. Code Rate

Code Rate can be defined as number of message bits / number of transmitted bits (k / n). As code rate decreases meaning the number of transmitted encoded bits increases, security or protection of message increases. But as code rate decreases bandwidth increases and also the cost per bit increases as total number of transmitted bits increases.

## II. REPRESENTATION OF LDPC CODES

Basically, these parity check sets can be represented in two ways. The first way is to describe in matrix form like all other linear block codes and second way is the graphical representation [3].

### A. Matrix and Algebraic Representation

In LDPC code, the parity check sets can be expressed in terms of Sparse Parity Check Matrix. In a Sparse Parity Check Matrix which is having (n-k) ✕ n dimension, the word "Sparse" means that the number of times '1's are very less than the number of times '0's. Normally for a large bit stream, the LDPC parity check matrix are of 1000✕2000 dimensions. So, out of n✕(n-k) entries, the number ones are very less than the number of zeros. There are three parameters that define the sparse parity check matrix which are (n, $w_c$, $w_r$). Here, n is coded length, $w_r$ is the number of ones in a row and $w_c$ is the number of ones in a column. For a matrix to be called low-density or sparse, the condition - $w_c$ and $w_r$ << n✕(n-k) be satisfied. To satisfy this, the parity check matrix must be very large.Here, a 4✕8 sparse parity check matrix H of (8,4,2) is shown.Each row of matrix H represents a check node and each column of matrix H represents variable nodes.

$$H = \begin{bmatrix} 0 & 1 & 0 & 1 & 1 & 0 & 0 & 1 \\ 1 & 1 & 1 & 0 & 0 & 1 & 0 & 0 \\ 0 & 0 & 1 & 0 & 0 & 1 & 1 & 1 \\ 1 & 0 & 0 & 1 & 1 & 0 & 1 & 0 \end{bmatrix}$$

Figure 1: Sparse Parity Check Matrix

Variable nodes indicate the elements of the code word. Check nodes indicate the set of parity check conditions. Here, as per the H matrix illustrated above, wherever there is a 1 in the matrix, it denotes that there is a connection between check node and variable node.

Parity Check Matrix can be classified into two types.

1) *Regular Parity Check Matrix:* If the parity check matrix has uniform $w_r$ and $w_c$(same no of ones in column and row) we call that a regular parity check matrix. This type makes decoding less complex.

2) *Irregular Parity Check Matrix:* If $w_c$ and $w_r$ are different for different columns and rows then it is called an Irregular parity check matrix.

B. *Tanner Graph Representation*

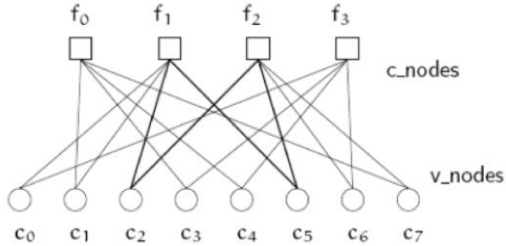

Figure 2: Tanner graph

Figure 2 is an example for such a tanner graph and represents the same code as the matrix from Figure 1. So, there will be n variable nodes and (n-k) check nodes. An edge or connection is made between the check node Ci and the variable node Vj if the element hij of matrix H is a 1. Check node $C_i$ is connected to variable node $V_j$ if the element $h_{ij}$ of matrix H is a 1. Since the rows in H matrix refers to the check nodes and columns of matrix H refers to the variable nodes, we can see that there are 4 check nodes .

III. LDPC ENCODING

There are various encoding and decoding schemes for LDPC codes. All schemes are there with the same aim of making the less complex and faster encoding and decoding process.
Most common approach for the encoding process based on Linear Block Code is described below.

A. *Linear Block Code Encoding*

We know that a codeword of Linear Block Code (n,k) is made up of the message bits and parity bits which we can describe as below .
$$C = |m_{1 \times k} : P_{1 \times n-k}|$$
where, p = parity vector

m = message vector

As per the property of Linear Block Code,
$$C . H^T = 0$$
To satisfy this condition, after matrix multiplication we get the equations for $P_{n-k \times 1}$ in terms of message bits. So, after getting the values of $P_{n-k \times 1}$, we can get the codeword of n bits.

B. *Generator Matrix Formation based Encoding*

With help of parity-check matrix H Generator matrix can be found by performing Gauss-Jordan elimination on H to obtain it in the form H = [A $I_{n-k}$], where A is a (n-k) × k binary matrix and In*k is the size of the (n-k) identity matrix. The generator matrix is then G = [ $I_K$ $A^T$ ]. And then by multiplying the message bit stream with the generator matrix will get us a codeword of n bits.
$$C = m . G$$
However, in this approach unlike the H matrix, the matrix G will most likely not be sparse. That will be a drawback of this method and since the matrix multiplication is quite big as there are thousands of bits, so the encoder becomes complex. For that a structured parity-check matrix formation has to be used to significantly lower the implementation complexity. But that does not go with the random selection of the parity check set we discussed earlier. So, a good approach is to avoid constructing a G matrix at all and instead to encode using other encoding schemes.

IV. LDPC DECODING

Decoding algorithms for LDPC codes have been discovered and formed several times independently and hence comes under different names. The most common ones are the bit flipping algorithm and the sum-product algorithm. Basically, decoding is divided into two types hard decision decoding and soft decision decoding. The class of decoding algorithms are collectively termed as message-passing algorithms since their operations can be explained by the passing of messages along the edges of a Tanner graph or we can say passing between variable and check nodes. As the encoded codewords are received then they are passed into this algorithm in which they pass back and forward between the variable and check nodes iteratively until a result is obtained which is erasure free. Thus, they are also known as iterative decoding algorithms.
IV.A.1
A. *Bit-flipping Decoding*

This is a hard-decision message-passing algorithm for LDPC codes. It is based on the principle that a codeword bit involved in a large number of incorrect check equations is likely to be incorrect itself. For every received bit a binary hard decision is made by the detector and is passed to the decoder. The messages passed along the Tanner graph edges are also binary: a bit node sends a message declaring if it is a one or a zero, and each check node sends a message to each connected bit node, declaring what value of the bit is based on the information available to the check node. The check node determines whether the parity-check equation is satisfied or not. If the modulo-2 sum of the incoming bit values is zero then the condition is satisfied. If the majority of the messages received by a bit node are different from its received value then the bit node flips (1 to 0 or vice versa) its current value. This process is repeated until all of the parity-check equations are satisfied. The bit-flipping decoder can be immediately stopped whenever a valid codeword has been found by checking if all of the parity-check equations are satisfied. This is valid for all message-passing decoding of LDPC codes and has two important benefits: firstly, additional iterations are avoided once a solution has been found, and secondly a failure to converge to a codeword is always detected.

### B. Sum-Product Decoding

The sum-product algorithm is a soft decision type message-passing algorithm. It is quite similar to the bit-flipping algorithm but instead messages that represent each decision in the bit-flipping algorithm, now represent probabilities. Bit-flipping decoding accepts an initial hard decision on the received bits as input whereas the sum-product algorithm is a soft decision algorithm which accepts the probability of each received bit as input.

## V. SIMULATION RESULTS

From the results it can be concluded that using LDPC codes lead to lower Bit Error Rate (BER) values. It can be noted that it is an iterative decoding process instead of Trellis based. Another advantage of LDPC codes is that the decoding complexity decreases with the length of the blocks. From the SNR vs BER graph, it is observed that at low SNR we can achieve very low BER compared to other coding schemes. And also, we can correct multiple erasures which occur during the transmission through the channel.

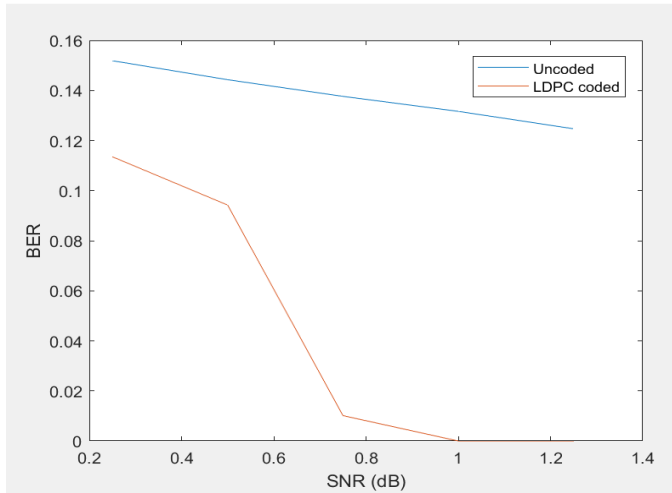

Figure 3: BER vs SNR curve for LDPC coded vs uncoded signal.

## VI. CONCLUSION

The reinvention of LDPC codes was an important event in the study of capacity-approaching codes. These are one of the interesting topics in coding theory. LPCD codes, unlike many others, are already equipped with very fast (probabilistic) encoding and decoding algorithms with the introduction of important variations of the sum product algorithm like logarithmic sum product. It is best suited for practical applications. LDPC codes are becoming the mainstream in communication generations. They are already implemented in 3G and LAN standards and are being implemented for 5G standards [4] . The other applications of LDPC codes are in video transmission like DVB-S2[5] and in satellite communication.

### ACKNOWLEDGEMENT

We are very grateful to Prof. Rachna Sharma, who provided us the chance to study and gather useful information on LDPC codes.